\documentclass[aps,prl,preprint,groupedaddress,showkeys]{revtex4}
\usepackage{graphicx,color}
\begin{document}

\preprint{ver1}

\title{Surface Majorana Cone of the Superfluid $^3$He B Phase
}


\author{Satoshi Murakawa}
\altaffiliation{Present address: Department of Physics, Keio University, Yokohama, Kanagawa 223-8522, Japan}
\author{Yuichiro Wada}
\author{Yuta Tamura}
\author{Masahiro Wasai}
\author{Masamichi Saitoh}
\altaffiliation{Present address: Institute of Physics and TIMS, University of Tsukuba, Tsukuba, Ibaraki 305-8571, Japan}
\author{Yuki Aoki}
\altaffiliation{Present address: Department of Materials Science and Engineering,
Tokyo Institute of Technology, Yokohama, Kanagawa 226-8502, Japan}
\author{Ryuji Nomura}%
\email[]{nomura@ap.titech.ac.jp}
\author{Yuichi Okuda}%
%\thanks{}
\affiliation{Department of Condensed Matter Physics, Tokyo Institute of Technology, 2-12-1 O-okayama, Meguro-ku, Tokyo 152-8551, Japan}

\author{Yasushi Nagato}
\affiliation{Information Media Center, Hiroshima University, Kagamiyama 1-4-2, Higashi-Hiroshima, 739-8511 Japan}
\author{Mikio Yamamoto}
\author{Seiji Higashitani}
%\affiliation{Graduate school of Integrated Arts and Sciences, Hiroshima University, Kagamiyama 1-7-1, Higashi-Hiroshima, 739-8511 Japan}
\author{Katsuhiko Nagai}
\affiliation{Graduate school of Integrated Arts and Sciences, Hiroshima University, Kagamiyama 1-7-1, Higashi-Hiroshima, 739-8511 Japan}
%\affiliation{Institute for Advanced Materials Research, Hiroshima University, Kagamiyama 1-3-1, Higashi-Hiroshima, 739-8521 Japan}


\date{\today}

\begin{abstract}
The superfluid $^3$He B phase, one of the oldest unconventional fermionic condensates experimentally realized, 
is recently predicted to support Majorana fermion surface states. 
Majorana fermion, which is characterized by the equivalence of particle and antiparticle, 
has a linear dispersion relation referred to as the Majorana cone. 
We measured the transverse acoustic impedance $Z$ of the superfluid$^3$He B phase changing its boundary condition 
and found a growth of peak in $Z$ on a higher specularity wall. 
Our theoretical analysis indicates that the variation of $Z$ is induced by
the formation of the cone-like dispersion relation and thus confirms the important feature of the Majorana fermion in the specular limit.
\end{abstract}

%\pacs{67.57.Np, 43.58.+z, 74.45.c}

\maketitle


Surface Andreev bound states (SABS) of the superfluid $^3$He B phase are
receiving renewed attention as ``edge states" of a 3D time reversal
invariant 
topological superfluid\cite{Schnyder,Roy,Kitaev,Qi,Chung,Volovik1,Volovik2,Nagato09,Machida}. 
Topological superfluids and superconductors are characterized by a non-trivial topological number in the gapped bulk state and 
a gapless edge state on their edges or surfaces. 
SABS  of the superfluid $^3$He B phase can be regarded as Majorana
fermions as they satisfy the Majorana condition, i.e., a particle and its antiparticle are equivalent, 
because the degrees of freedom of the bound states are halved.
When a surface is specular, 
linear dispersions are predicted for Majorana fermions\cite{Nagato1}, which have been recently referred to as a Majorana cone\cite{Qi,Chung}.
We measured the transverse acoustic impedance $Z$ of the superfluid
$^3$He B phase changing its boundary condition up to practically
specular scattering.
The observed variation of $Z$ is well reproduced by theoretical
analysis and is shown to be induced by
the formation of the cone-like dispersion relation of SABS at higher
specularities. This provides an evidence of the existence of the
Majorana cone
in superfluid ${}^3$He B in the specular limit.

\begin{figure}[h]
	\includegraphics[width = 5.5 cm]{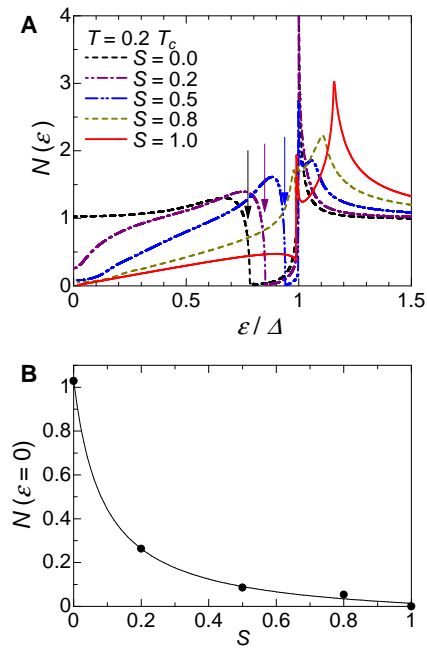}%
	\caption{(Color online) (A) Theoretically calculated surface density of states for the superfluid $^3$He B phase 
as a function of the quasiparticle energy $\varepsilon$ at $T=0.2T_c$ and various specularities $S$. 
Vertical axis is normalized to the normal state value and 
the horizontal axis is normalized to the superfluid gap $\Delta$.
Arrows represent the band edge energy $\Delta^*$ of the SABS. 
(B) Zero-energy weight of the surface density of states as a function of $S$. 
Line is to guide the eye.    
		}
\end{figure}
\begin{figure}[t]
	\includegraphics[width = 5.5 cm]{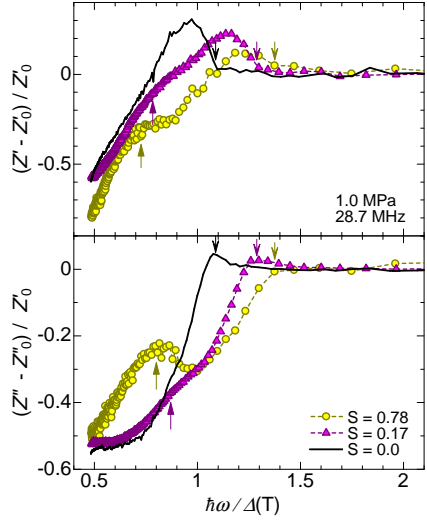}%
	\caption{(Color online) Experimentally observed transverse acoustic impedance $Z$ as a function of the acoustic energy scaled to the superfluid gap $\Delta(T)$. 
Typical data taken at a frequency $\omega/2\pi=28.7$ MHz and a pressure $P=1.0$ MPa are shown here, 
but  similar behaviors are observed at other $\omega$ and $P$. 
Real $Z'$ and imaginary $Z''$ components are shown in the upper and lower panel separately 
at specularities $S=$ 0, 0.17$\pm0.1$, and 0.78$\pm0.1$. $Z_{0}$ is the normal state value at each $S$ just above the transition temperature. 
Downward arrows indicate the high-energy (or high-temperature) singularities. 
Low-energy (or low-temperature) peaks denoted by the upward arrows grow at higher $S$. 
		}
\end{figure}
\begin{figure}[h]
	\includegraphics[width = 5.5 cm]{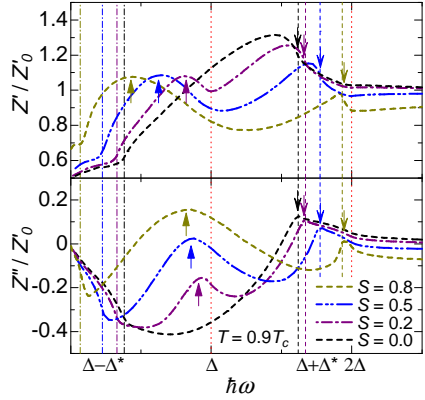}%
	\caption{(Color online) Theoretically calculated transverse acoustic impedance $Z$ as a function of the acoustic energy $\hbar\omega$ 
at specularities $S=$  0, 0.2, 0.5, and 0.8 at $T=0.9T_c$. 
Real $Z'$ and imaginary $Z''$ components scaled to the normal state value $Z'_{0}$ are shown in the upper and lower panel, respectively. 
Vertical dotted lines are the characteristic energies of the superfluid gap $\Delta$ and $2\Delta$. 
Vertical dashed and dot-dashed lines are $\Delta+\Delta^*$ and $\Delta-\Delta^*$, respectively. 
$\Delta^*$ is the band edge energy of the SABS. 
Downward arrows indicate the high-energy singularities at $\Delta+\Delta^*$. 
Low-energy peaks are denoted by the upward arrows, which become larger at higher $S$ due to formation of the Majorana cone.  
		}
\end{figure} 

\begin{figure*}[b]
	\includegraphics[width = 15 cm]{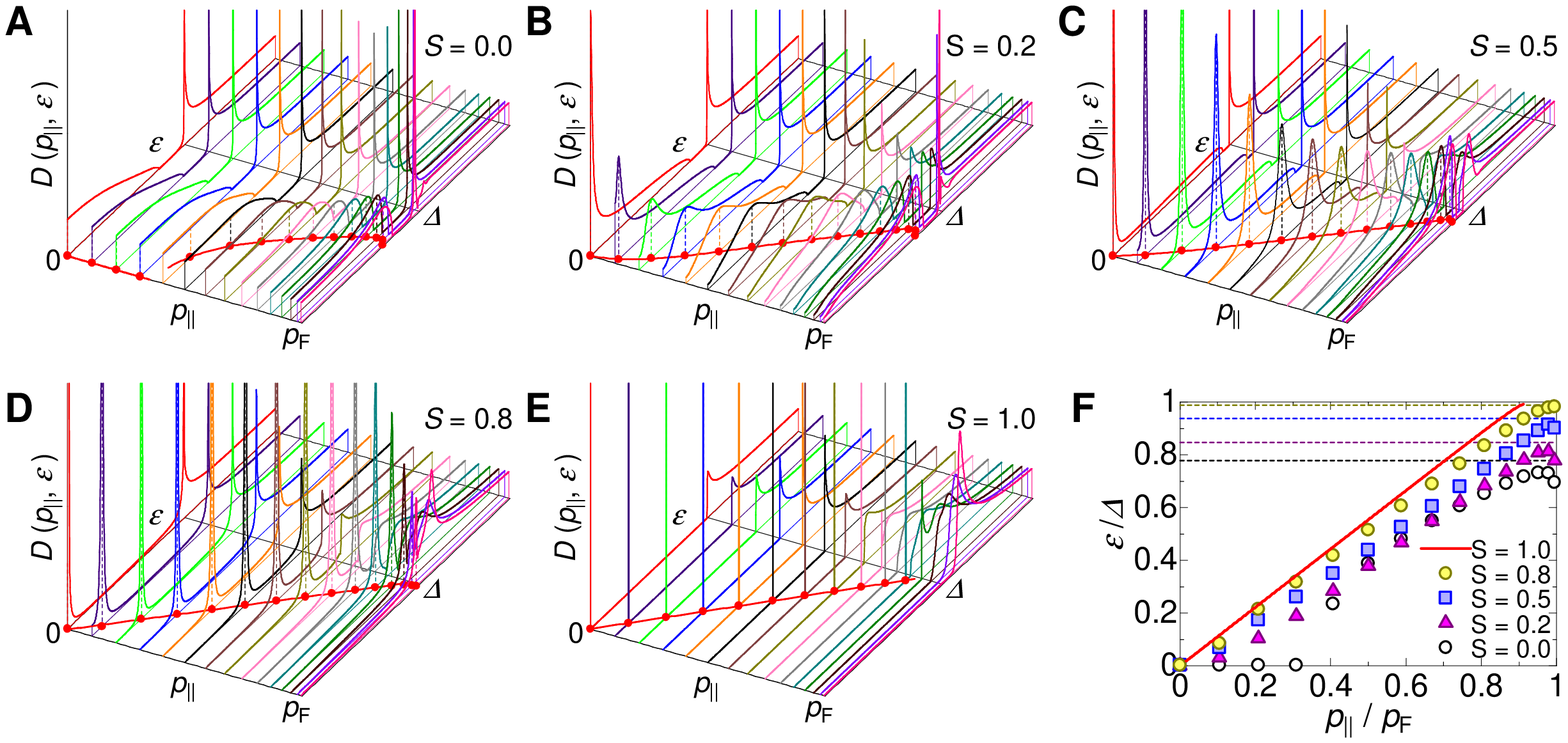}%
	\caption{(Color online) (A)-(E) Theoretically calculated angle resolved surface density of states of the superfluid $^3$He B phase 
as functions of the parallel component of the momentum $p_{\parallel}$ and the quasiparticle energy $\varepsilon$ 
at specularities $S=$ 0, 0.2, 0.5, 0.8, and 1. 
Vertical axis is normalized to the normal state value and $p_{F}$ is the Fermi momentum. 
Although the density of states is a continuous function of $p_{\parallel}$, it is plotted discretely for clarity.  
Results at low temperature $0.2 T_c$ are shown here. 
Temperature effect is weak and dispersions at higher temperatures appear to be similar 
if the energy is scaled to the superfluid gap $\Delta$ at the temperatures. 
Circles are the peak positions of the spectrum, and are shown altogether in (F). 
Horizontal dashed lines  in (F) are the band edge energies $\Delta^*$. 
		}
\end{figure*}

Superfluid $^3$He is a well established spin-triplet p-wave superfluid 
where the B phase is the realization of the Balian-Werthamer state, which breaks the relative spin-orbit symmetry\cite{Vorhalt,Halperin}. 
The surface Majorana states of the $^3$He B phase should exhibit peculiar features such as a surface spin current 
and an anisotropic magnetic response\cite{Chung,Volovik2,Nagato09,Machida}.  
Under the specular scattering boundary condition of the surface, 
SABS with a finite energy are not degenerate, and have a dispersion relation linear to the momentum, which takes the shape of a 2D cone
for massless Dirac fermions.
In the spin-triplet p-wave pairing system, the annihilation operator of
the
negative energy state is equivalent to the creation operator of the
positive energy state. Hence, only the positive energy
cone, which is called a Majorana cone, is physical\cite{Qi,Chung}. 
To date, Majorana surface fermion states have been discussed mostly on smooth surfaces (specular scattering limit). 
However, intriguing problems are to elucidate how the Majorana cone transforms as the roughness of the walls increases 
(or the specularity $S$ decreases) 
and how this transformation affects the experimental observations.  

$S$ is a parameter introduced to specify the boundary condition for quasiparticle scattering. 
$S=1$ corresponds to the specular scattering limit where quasiparticles conserve the parallel component of the momentum and 
only the perpendicular component is inverted after scattering off a wall. 
In contrast, $S=0$ is the diffusive limit where quasiparticles are scattered in random directions. 
For a partially specular scattering condition $0<S<1$, a fraction $S$ of the quasiparticles is specularly
scattered, while the remaining fraction $1-S$  is
scattered diffusely\cite{Milliken,Einzel,Nagato1}. In liquid ${}^3$He, $S$ can be controlled {\it in situ} by coating the wall with thin layers of $^4$He\cite{Tholen,Kim}. 
$^4$He atoms are selectively adsorbed onto the walls due to their larger mass and reduce the roughness of the wall.

Figure 1A shows a theoretical calculation\cite{Nagato1,Nagato2,Murakawa}
 for the angle averaged 
surface density of states (SDOS)
$N(\varepsilon)$
for the B phase at various $S$. 
In the diffusive limit where $S=0$, a nearly flat SABS band with a very
sharp edge at $\Delta^*$ appears below the bulk superfluid gap $\Delta$. The
existence of $\Delta^*$ was theoretically predicted many years ago\cite{Zhang2,
Nagato1,Vorontsov}. 
As $S$ increases, the edge $\Delta^*$ increases and the low-energy spectral weight shifts toward the band edge 
forming a peak structure just below $\Delta^*$.
For the specular limit where $S=1$, SDOS increases linearly with energy at low energy: $N\propto\varepsilon$. 
This linear energy dependence reflects the linear dispersion of
SABS or the Majorana cone. As is shown in Fig. 1B, the density of
states at zero energy $N(\varepsilon=0)$ decreases rapidly as $S$ increases.

We measured the temperature dependence 
of the transverse acoustic impedance $Z(T)$ of the superfluid $^3$He B
phase at various $S$ values
using AC-cut quartz transducers immersed in liquid $^3$He. 
In this case, the wall corresponds to the surface of the transducer. 
$Z=Z'+iZ''$ is a complex quantity defined as the ratio of the shear
stress $\Pi$ of the liquid at the wall to
the oscillation velocity $u$ 
\begin{equation}
Z=\frac{\Pi}{u}. 
\end{equation}
$Z$ can be obtained from the change in the resonance angular frequency
${\Delta}\omega$ and the quality factor ${\Delta}Q$ 
of the quartz transducers, 
\begin{equation}
    Z = Z'+iZ'' = \frac{1}{4}n\pi Z_q\Delta Q^{-1}
+i\frac12 n\pi Z_q \frac{\Delta \omega}{\omega}.
\end{equation}
Here $\omega$ is the measured angular frequency, $Z_q$ is the acoustic impedance of the quartz transducer
and $n$ is the harmonics number\cite{Halperin}. 
$S$ is determined independently by measuring $Z(T)$ in a normal fluid
with ${}^4$He coating. 
The temperature dependence of $Z$ is largest for pure
$^3$He without coating, where the boundary condition is supposed to be 
at the diffusive limit $S=0$.
The temperature dependence becomes smaller in the sample with $^4$He coating
due to the reduced transverse momentum exchange 
and $Z(T)=0$ at $S=1$. 
$S$ can be obtained by fitting $Z(T)$ with the Fermi liquid theory\cite{Milliken,Einzel}.  
Details of the experimental procedure have already been reported\cite{Wada}.

We have demonstrated in earlier publications that $Z$ is a unique surface probe for SABS in superfluid $^3$He 
and have provided the spectroscopic details of SDOS\cite{Aoki,Saitoh,Wada,Murakawa}. 
Figure 2 presents 
the measured $Z(T)$ as a function of the normalized energy 
$\hbar\omega/\Delta(T)$.
$\Delta(T)$ is the superfluid gap energy at $T$ obtained from the weak-coupling-plus model\cite{Halperin}; 
thus, a higher (lower) energy corresponds to a higher (lower) temperature.
$Z_{0}$ is the normal state value of $Z$ just above the transition temperature $T_c$. 
To view data at various $S$ on the same scale, 
the vertical axis is the difference from the normal state value $Z-Z_{0}$ scaled by $Z'_{0}$. 

In a pure ${}^3$He system without ${}^4$He coating, the boundary
condition
is supposed to be fully diffusive, i.e., $S=0$.
In this system, we found a kink in the real part $Z'$ of the impedance
and a peak in the imaginary part $Z''$ at a higher frequency, 
which are indicated by the downward arrows in Fig. 2\cite{Aoki, Saitoh}. 
Theoretical analysis showed that
these singularities are due to pair excitations of a SABS quasiparticle 
and a propagating Bogolubov quasiparticle\cite{Nagato2} with  frequency
$\hbar\omega=\Delta(T)+\Delta^*$. In fact, the kink structure in the
real part $Z'$ is well
reproduced by a joint density of states
\begin{equation}
 \int_0^{\hbar\omega}d\varepsilon N(\varepsilon)N(\hbar\omega-\varepsilon),
\end{equation} 
where $N(\varepsilon)$ is the SDOS shown in Fig. 1.
In the system with ${}^4$He coating, $\Delta^*$
increases
in accordance with theory\cite{Wada}.

In partially specular cases, new peaks grow in the low-energy region
as $S$ increases\cite{Murakawa}, which are indicated by the upward arrows in Fig. 2.
At the highest specularity $S=0.78\pm0.1$ we have achieved to date, the peaks
become very distinctive. 
The theoretical calculation reproduces the growth of the low-energy peak at higher $S$. 
Figure 3 shows the theoretical results of the frequency dependence of $Z$ for various
values of $S$.
The calculation was performed in the same manner as 
Refs. \onlinecite{Nagato2} and \onlinecite{Nagai}.

Let us consider the origin of the new peak in partially specular cases.
We show in Fig. 4A-E 
the angle resolved SDOS at various $S$.
The spectra are plotted as functions of the parallel component of the momentum $p_{\parallel}=p_{F}{\sin}\theta$ 
where $\theta$ is the incident angle of the quasiparticle measured from the surface normal and $p_{F}$ is the Fermi momentum. 
The circles in Figs. 4A-E indicate the peak positions in the spectra, and are plotted altogether in Fig. 4F. 
Although the peak position is insignificant due to the very broad spectrum at $S=0$, 
we plotted the maximum positions at every $S$. 
At $S=0$ (Fig. 4A), states below $\Delta^*$ appear for all incident angles\cite{Nagato1,Nagato2,Zhang2,Vorontsov} 
and a cone-like structure is not recognized in the broad spectrum.
However, even at a relatively small specularity $S=0.2$ (Fig. 4B),  
the zero-energy weight is significantly suppressed, except around the normal incidence $p_{\parallel}\, {\approx} \, 0$. 
A cone-like structure is formed in this case, but the peak positions 
are still below the linear dependence, as found in Fig. 4F. 
At $S=0.5$ (Fig. 4C), which is within the range of the experiment, 
a rather sharp cone is recognized and the peak positions follow a nearly linear dependence $\varepsilon=cp_{\parallel}$ 
with $c \, {\approx} \, \Delta/p_{F} \, {\approx} \, 5$ cm/s  
as in the specular limit $S=1$ (Fig. 4F). 
Thus, the Majorana cone like structure is realized at higher specularities.

It was shown that $Z$ in low-energy region is dominated 
by the scattering of the thermally excited quasiparticles\cite{Murakawa}. 
The low-energy peak in $Z$ begins to grow at $\hbar\omega=\Delta-\Delta^*$ (dot-dashed lines 
in Fig. 3) as $\hbar\omega$ increases. 
Here, $\Delta-\Delta^*$ corresponds to the energy difference between the band edge and the propagating Bogoliubov states with $\varepsilon \ge \Delta$. 
The growth in $Z$ is a clear sign of the inelastic scattering of quasiparticles
from the thermally occupied SABS around the SDOS maximum to the propagating Bogoliubov states.
$Z$ falls around $\hbar\omega=\Delta$,
since the low-energy SDOS is suppressed by the formation of the
Majorana cone. This is in contrast to the diffusive limit ($S=0$ in Fig. 3).
Since the SDOS in the diffusive limit remains finite down to zero energy,
no double peak structure is found.
The growth of the low-energy peak is a distinct evidence of the formation of the surface Majorana cone.

In summary, the growth of low-energy peaks in the transverse acoustic impedance is observed 
in a topological superfluid, the superfluid $^3$He-B phase. 
This growth is reproduced theoretically by a self-consistent calculation.
The dispersion of SABS is calculated for various boundary conditions and 
a cone-like structure is clearly observed even on a partially specular wall. 
The experimentally observed growth of the peaks is due to 
the formation of a surface Majorana cone on a higher specularity wall, 
and is a strong indicator of the predicted Majorana cone in the specular limit. 
Transverse acoustic impedance measurement is a promising technique for studying surface states and 
will provide more direct proof of the Majorana nature as the anisotropic magnetic response of the Majorana cone\cite{Chung,Nagato09}.    

\section{Acknowledgements}
This study was supported in part by
the Global COE Program through the Nanoscience and
Quantum Physics Project of Tokyo Tech. and by 
a Grant-in-Aid for Scientific Research
on Priority Areas ``Physics of Super-clean Materials'' (No. 20029009 and 17071009) 
and on Innovative Areas ``Topological Quantum Phenomena'' (No. 22103003) 
from MEXT of Japan.

%Create the reference section using BibTeX:

%\bibliographystyle{apsrev}
%\bibliography{nagai}

\begin{thebibliography}{00}
\bibitem{Schnyder}  A. P. Schnyder, S. Ryu, A. Furusaki, and A. W. W. Ludwig: 
Phys. Rev. B {\bf 78} (2008) 195125.
\bibitem{Roy} R. Roy:  
arXiv:0803.2868v1 (2008).
\bibitem{Kitaev} A. Kitaev:  
AIP Conference Proceedings, {\bf 1134} (2009) 22.
\bibitem{Qi} X. L. Qi, T. L. Hughes, S. Raghu, and S. C. Zhang:  
Phys. Rev. Lett. {\bf 102} (2009) 187001.
\bibitem{Chung} S. B. Chung and S. C. Zhang:  
Phys. Rev. Lett. {\bf 103} (2009) 235301.
\bibitem{Volovik1} G. E. Volovik: 
JETP Lett. {\bf 90} (2009) 587.
\bibitem{Volovik2}  G. E. Volovik:
JETP Lett. {\bf 90} (2009) 398.
\bibitem{Nagato09} Y. Nagato, S. Higashitani, and K. Nagai:
J. Phys. Soc. Jpn. {\bf 78} (2009) 123603.
\bibitem{Machida} Y. Tsutsumi, M. Ichioka, and K. Machida: arXiv:1010.3066 (2010).
%\bibitem{Nayak} Nayak, C., Simon, S. H., Stern, A., Freedman, M. \& Sarma, S. D. 
%Non-Abelian anyons and topological quantum computation. 
%{\it Rev. Mod. Phys.} {\bf 80}, 1083-1159 (2008). 

\bibitem{Nagato1} Y. Nagato, M. Yamamoto, and K. Nagai:
J. Low Temp. Phys. {\bf 110} (1998) 1135.


\bibitem{Vorhalt}  D. Vollhardt and P. W\"{o}lfle:
The Superfluid Phases of Helium 3 (Taylor and Francis, London, 1990).
\bibitem{Halperin} W. P. Halperin and E. Varoquaux:
Helium Three (eds W. P. Halperin and L. P. Pitaevskii) (Elsevier, Amsterdam, 1990). 


\bibitem{Milliken} F. P. Milliken, R. W. Richardson, and S. J. Williamson: 
J. Low Temp. Phys. {\bf 45} (1981) 409.
%\bibitem{Richardson} R. W. Richardson, Phys. Rev. B {\bf 18}, 6122 (1978).
\bibitem{Einzel}  D. Einzel, P. W\"{o}lfle, and P. J. Hirschfeld: 
J. Low Temp. Phys. {\bf 80} (1998) 31.

\bibitem{Tholen} S. M. Tholen and J. M. Parpia: 
Phys. Rev B {\bf 47} (1993) 319.
\bibitem{Kim} D. Kim,  M. Nakagawa, O. Ishikawa, T. Hata, and T. Kodama:
Phys. Rev. Lett. {\bf 71} (1993) 1581.



\bibitem{Murakawa} S. Murakawa, Y. Tamura, Y. Wada, M. Wasai, M. Saitoh, Y. Aoki, R. Nomura, Y. Okuda, 
Y. Nagato, M. Yamamoto, S. Higashitani, and K. Nagai: 
Phys. Rev. Lett. {\bf 103} (2009) 155301.
\bibitem{Nagato2} Y. Nagato, M. Yamamoto, S. Higashitani, and K. Nagai:
J. Low Temp. Phys. {\bf 149} (2007) 294.

\bibitem{Zhang2} W. Zhang:
Phys. Lett. A {\bf 130} (1988) 314.
\bibitem{Vorontsov} A. B. Vorontsov and J. A. Sauls:  
Phys. Rev. B {\bf 68} (2003) 064508.

\bibitem{Wada} Y. Wada, S. Murakawa, Y. Tamura, M. Saitoh, Y. Aoki, R. Nomura, and Y. Okuda: 
Phys. Rev. B {\bf 78} (2008) 214516.
\bibitem{Aoki} Y. Aoki, Y. Wada, M. Saitoh, R. Nomura, Y. Okuda, Y. Nagato, M. Yamamoto, S. Higashitani, and K. Nagai: 
Phys. Rev. Lett. {\bf 95} (2005) 075301.
\bibitem{Saitoh} M. Saitoh, Y. Wada, Y. Aoki, S. Murakawa, R. Nomura, and Y. Okuda:   
Phys. Rev. B {\bf 74} (2006) 220505(R).


\bibitem{Nagai} K. Nagai, Y. Nagato, M. Yamamoto, and S. Higashitani: 
J. Phys. Soc. Jpn. {\bf 77}, (2008) 111003.



\end{thebibliography}

\end{document}